# Electrically Modulated Wetting of Drops on Soft Dielectric Films

Ranabir Dey[1], Sunando DasGupta[2,3], Suman Chakraborty[1,3]*

[1]*Department of Mechanical Engineering, Indian Institute of Technology Kharagpur, Kharagpur- 721 302, West Bengal, India.*

[2]*Department of Chemical Engineering, Indian Institute of Technology Kharagpur, Kharagpur- 721 302, West Bengal, India.*

[3]*Advanced Technology Development Centre, Indian Institute of Technology Kharagpur, Kharagpur- 721 302, West Bengal, India.*

---

* Corresponding author, E-mail: suman@mech.iitkgp.ernet.in

**Abstract**

The inter-connection between the elasticity of a dielectric film and the wetting of a sessile drop on the same, under an applied electrical voltage, remains unaddressed. Here, we report the electrowetting-on-dielectric (EWOD) behaviour of sessile drops on dielectric elastomer films of varying elasticities- from an apparently rigid dielectric film to a soft, deformable dielectric film. Our results reveal that the elasticity of the underlying dielectric film provides an additional control over the droplet electrowetting behaviour, which may be best addressed from free energy based consideration, leading to a modification of the classical Lippman-Young paradigm. We also provide an explanation on the displacement profiles for the deformation of the soft dielectric surface, due to the interfacial electro-elastocapillary interaction triggered by the electrowetted sessile droplet. These results can be of profound importance in various emerging applications, ranging from the development of soft liquid lenses to drug delivery.

## 1 Introduction

Application of an electrical voltage, to a conducting sessile droplet resting on a thin dielectric film covering a planar electrode, results in enhanced wetting of the droplet. Such electrically controlled wetting of a droplet, on a dielectric film, is classically referred to as 'electrowetting-on-dielectric (EWOD)' or 'electrowetting'[1,2]. The detailed understanding of, and control over, the electrowetting phenomenon, developed over the years, have translated into several state-of-the-art applications, such as variable focus lenses[2,3], reflective displays[2,3], microfluidic mixing[4], control of droplet morphology over functional substrates[5], control of droplet detachment from hydrophobic surfaces for the development of three-dimensional digital microfluidic systems[6], thermal management of miniaturized devices[7], and several bio-medical applications[8]. Relatively recently, EWOD of a sessile droplet was used for controlling the folding and unfolding of a thin membrane around a liquid droplet[9]. Such electrically controlled reversible wrapping of a thin membrane forms the foundation for electrically assisted capillary origami[10].

Despite these emerging applications, the physical understanding of droplet electrowetting, at least for practical purposes, is still largely limited to the Lippmann-Young equation[1,2]. One of the major shortcomings of the Lippmann-Young equation is that it is intrinsically restricted to rigid dielectric films. However, the electrowetting on soft substrates may lead to intricate electrically mediated elastocapillary interactions, or electro-elastocapillary interactions, which are not intrinsically captured through this classical paradigm.

Here, we first experimentally investigate the electrowetting behaviour of sessile drops on dielectric elastomer films of varying elasticities. Specifically, we try to bring out the differences between the electrowetting characteristics of a conducting sessile drop on an apparently rigid dielectric elastomer film, and those on a soft dielectric elastomer film. We show that the droplet electrowetting on the soft dielectric elastomer film cannot be addressed by the classical Lippmann-Young equation. Thereafter, we try to explain the electrowetting behaviour on the soft, deformable dielectric film by a free energy minimization approach. Our analysis takes into consideration that the reduction in elasticity of the dielectric elastomer film incurs an additional elastic energy, due to the surface deformation of the soft dielectric film by the electrowetted sessile droplet; it also considers an alteration in the electrostatic energy, due to the effective coupling between the dielectric constant and the elasticity of the dielectric film for a general dielectric elastomer. We further provide here a detailed discussion on the displacement profiles for the deformation of the soft dielectric surface, due to the interfacial electro-elastocapillary interaction triggered by the electrowetted sessile droplet. Our results are likely to have profound implications towards the understanding of several previously inconceivable engineering and bio-medical applications, involving the interplay of electrowetting and substrate-compliance over solid-liquid interfacial scales.

## 2 Experimental section

### 2.1 Fabrication of the electrode-and-dielectric platform for the electrowetting-on-dielectric (EWOD) experiments

For the EWOD experiments, the dielectric films of different elasticities are fabricated from Sylgard 184 (*Dow Corning, USA*) - a Polydimethylsiloxane (PDMS) based elastomer. Sylgard 184 comprises of two components: A base monomer and a curing agent or a cross-linker. The dielectric films are fabricated from Sylgard 184 prepared by mixing the base and the cross-linker in the weight ratios of 10:1 and 50:1, followed by subsequent curing. Bulk rheometry test results, reported in the established literature, reveal that the Young's modulus $(E)$ of the differently cross-linked Sylgard 184 films decreases from 1.5 MPa to 0.02 MPa with increasing base-to-cross-linker ratio from 10:1 to 50:1[11–13] (see Table 1). In essence, the dielectric elastomer film becomes progressively 'softer' (i.e. $E$ decreases) with increasing base-to-cross-linker ratio. It must be noted here that for the present study, only the two dielectric films - 10:1 and 50:1 Sylgard 184 films, are chosen. This is done, because, based on the study of electrically triggered droplet spreading studies reported elsewhere[14], it can be concluded that the 10:1 Sylgard 184 film ($E = 1.5$ MPa) behaves as an apparently rigid dielectric film, on which the classical droplet electrospreading characteristics are valid; while the 50:1 Sylgard 184 film ($E = 0.02$ MPa) behaves as a soft, deformable one. Hence, the various aspects of electrowetting on soft dielectric films that we want to discuss in the present paper can be systematically highlighted by a comparative study between the electrowetting and the film surface deformation characteristics for the 10:1 Sylgard 184 dielectric film (apparently rigid) and those for the 50:1 Sylgard 184 dielectric film (soft). The different Sylgard 184 films, with tunable softness, are coated on glass slides, with transparent Indium Tin Oxide (ITO) film as the electrode layer (*Sigma Aldrich, surface resistivity: 70-100 Ω/sq*).

First, the square ITO coated glass slide is thoroughly cleaned by ultrasonication with acetone (*Merck*) for 10 minutes, followed by further ultrasonication with Milli-Q ultrapure water (*Millipore India Pvt. Ltd.*) for another 10 minutes. Thereafter, the ITO coated glass slide is dried in a nitrogen stream, and a small portion of the ITO coating is covered with a strip of parafilm tape (*Pachiney Plastic Packaging, USA*). The degassed, uncured Sylgard 184 is then poured carefully over the exposed part of the ITO coating on the glass slide. Thereafter, the elastomer is spin-coated (*spin coater: Süss MicroTec, Germany*) following a two-step procedure: In the first step, the spin velocity is 500 rpm for 30 seconds, while in the second step the spin velocity is maintained at 5000 rpm for 70 seconds, with an intermediate acceleration of 4000 rpm/s$^2$ and a final deceleration of identical magnitude. After coating the film, the parafilm strip is removed to reveal a thin strip of uncoated ITO underneath. The Sylgard 184 film is subsequently cured overnight at 95 °C. Identical procedure is followed for fabricating the dielectric films from Sylgard 184 with different base-to-cross-linker ratios (i.e. 10:1, and 50:1). The resulting thickness $(h)$ of each of the dielectric films, as measured by a surface profilometer (*Vecco Dektat 150*), is mentioned in Table 1. Furthermore, the surfaces of the dielectric films are characterized by evaluating the root-mean-square surface roughness $(r_{rms})$, by atomic force microscopy technique. The $r_{rms}$ value of each of the thin dielectric film surfaces is around 0.303 nm; hence, it is concluded that the surface roughness does not alter significantly with the dielectric film elasticity.

**Table 1:** Characteristics of the different dielectric elastomer films

|  | Sylgard 184 (10:1) | Sylgard 184 (50:1) |
|---|---|---|
| $E$ (MPa)[11–13] | 1.5 | 0.02 |
| $h$ (μm) | 9.65 | 9.43 |
| $\theta_{eq}^0$ (°) | $109.24 \pm 0.05$ | $112.41 \pm 2.25$ |
| $\theta_a^0 / \theta_r^0$ (°) | 115.6/102.7 | 126.95/29.06 |

## 2.2 *Experimental setup and procedure*

The electrowetting experiments are performed with sessile drops of 100 mM Potassium chloride (KCl) solution in Milli-Q ultrapure water (*Millipore India Pvt. Ltd.*). The physical properties of the 100 mM KCl solution are as follows: Surface tension, $\gamma = 72.75 \times 10^{-3}$ N/m, viscosity at room temperature of 20 °C, $\mu = 1.002 \times 10^{-3}$ Pa·s, and electrical conductivity, $\sigma_e = 1.215$ S/m. Moreover, the different dielectric films are also characterized by evaluating the equilibrium contact angle $\left(\theta_{eq}^0\right)$, the macroscopic advancing contact angle $\left(\theta_a^0\right)$, and the macroscopic receding contact angle $\left(\theta_r^0\right)$ of 100 mM KCl solution droplets on these films, without any electrical effects (see Table 1).

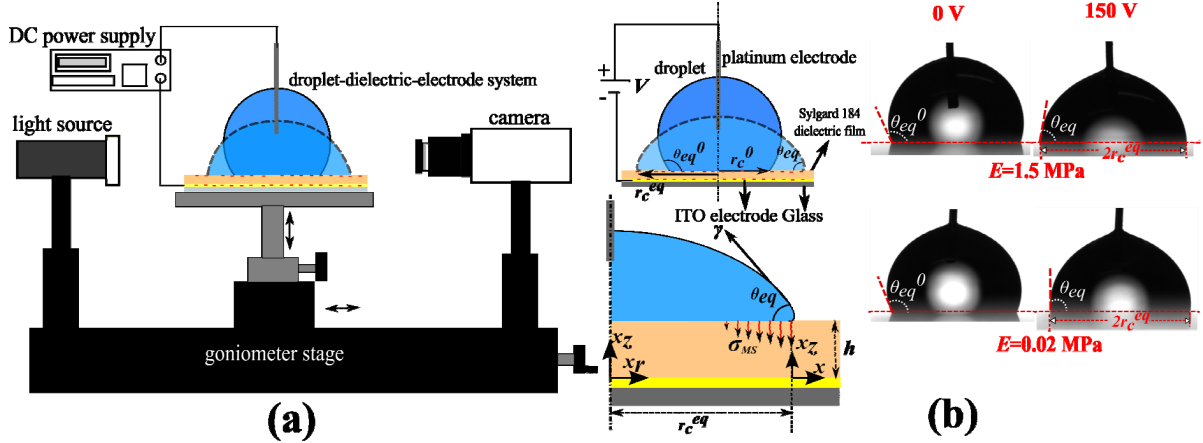

**Fig. 1: (a)** Schematic of the setup used for performing the electrowetting-on-dielectric (EWOD) experiments with dielectric elastomer films of different elasticity. **(b)** Schematic of the droplet-dielectric-electrode system used for the EWOD experiments. A blow-up schematic of the axisymmetric droplet wetting, under an applied electrical voltage, is shown here along with the co-ordinate systems used in the theoretical formulation. Sequential images, captured by the goniometer camera, exhibiting the profiles for the 100 mM KCl solution droplet before (0 V) and after application of an external electrical voltage (150 V), on the dielectric films of different elasticity ($E = 1.5$ MPa and $E = 0.02$ MPa), are also shown here. Schematics shown here are not to scale.

For determining the electrowetting characteristics, the ITO glass slide coated with a dielectric film, having a definite value of $E$, is first mounted on the platform of a Goniometer (*Ramé-hart instrument co., model no. 290-G1*) (see Fig. 1(a)). Then a sessile droplet of 100 mM KCl solution ($5 \pm 1$ μl in volume) is dispensed onto the dielectric film by a calibrated micro-syringe. The ITO electrode is electrically grounded by making electrical connections to the uncoated portion of the ITO. The electrical circuit is completed by means of a platinum wire electrode (dia: 160 μm) immersed into the droplet. The configuration of the electrowetting experimental setup, as discussed here, is schematically depicted in Fig. 1(a) and Fig. 1(b). The desired electrical voltage $(V)$ is applied between the ITO electrode and the platinum wire electrode by means of a DC sourcemeter (*Keithley 2410, 1100 V SourceMeter*). The applied DC electrical voltage is progressively increased from 0 to 150 V.

The maximum applied voltage $(V_{max} = 150 \text{ V})$ is fixed here by the fact that beyond $V \sim 150 \text{ V}$, lateral movement of the droplet is observed, and no repeatable experimental data could be obtained. Such observation is quite common in the electrowetting literature[15]. Moreover, beyond this electrical voltage limit, dielectric breakdown is also observed. On application of the electrical voltage, the sessile droplet spreads till the final steady-state is reached (see the schematic in Fig. 1(b)). At a definite value of $V$, the final macroscopic equilibrium droplet contact angle $(\theta_{eq})$ on the dielectric film and the final equilibrium droplet contact radius $(r_c^{eq})$ are measured using the Goniometer and the *Drop Image Advanced v2.2.3* software (see Fig. 1(b)). Sequential images exhibiting electrically controlled droplet wetting on dielectric films of different elasticity, as captured by the goniometer, are also shown in Fig. 1(b). Furthermore, at a definite value of $V$, the final contact radius of the electrowetted sessile droplet on the dielectric film is also evaluated using a self-written MATLAB (version 7.1; *The MathWorks Inc., USA*) image processing code. For a definite magnitude of the applied electrical voltage, the values of $r_c^{eq}$ obtained using the *Drop Image Advanced v2.2.3* software and the self-written image processing code are comparable.

## 3 Theoretical formulation

In this section, we discuss about the additional contributions to, and possible alterations of, the total free energy of the system in EWOD configuration (see the blow-up schematic in Fig. 1(b)) due to the reduction of the elasticity of the dielectric elastomer film from the 'apparently rigid' domain to the 'soft' domain.

In case of an electrowetted sessile droplet on a soft, deformable dielectric film, the total free energy $(F)$ of the EWOD system must generally include the additional elastic energy $(E_{el})$ stored due to the inherent film surface deformation. The evaluation of $E_{el}$ in turn involves the determination of the displacement profiles for the soft dielectric film surface deformation, due to the electrowetted sessile droplet on top of it. In this regard, it must be noted here that in case of an electrowetted sessile droplet, the underlying dielectric film surface deforms due to the combined interplay of the elastocapillary[11,16–19] interactions at the dielectric-droplet interface and the normal stress distribution on the film surface, due to the electrical stress (Maxwell stress) distribution at the solid-liquid interface. Another influence of the variation of the elasticity of the dielectric film on $F$ stems from the associated variation of the dielectric constant $(\epsilon_r)$ of the dielectric film with the varying value of $E$. In general for a dielectric elastomer, like Sylgard 184 (which is a Polydimethylsiloxane (PDMS) based elastomer), elasticity of the dielectric elastomer film is generally reduced by increasing the base polymer-to-cross-linker weight ratio[12,13,20,21]. The cross-linker (i.e. the curing agent) weight percentage decreases with the increasing base-to-cross-linker weight ratio. Consequently, the degree of polymerization of the monomeric units, i.e. the siloxane backbone in case of Sylgard 184, decreases with decreasing weight percentage of the cross-linker. Interestingly, the dielectric constant of the elastomer generally decreases with such decreasing degree of polymerization of the polymeric chain[22,23]. In essence, as the degree of polymerization decreases with increasing base polymer-to-cross-linker weight ratio, the Young's modulus of the dielectric elastomer film decreases, and so does the associated dielectric constant of the film. Hence, it can be said that $\epsilon_r$ is effectively a function of $E$ for dielectric elastomer films, and the former can be generally represented as $\epsilon_r(E)$. It must be noted here that the variation of the dielectric constant, with the varying higher values of the degree of polymerization, is insignificant[23]. Hence, the variation in $\epsilon_r$ with the varying higher values of $E$, which corresponds to the domain of higher values of degree of polymerization, is trivially neglected; accordingly, an average value of $\epsilon_r = 2.65$ is usually considered for apparently rigid Sylgard 184 films (e.g. 10:1 Sylgard 184 film). However, as $E$ is reduced here from the 'apparently rigid' to the 'soft' domain, where the latter corresponds to the domain of lower values of degree of polymerization, the corresponding sharp variation in $\epsilon_r$ cannot be trivially neglected[23]. Consequently, the associated variation in the electrostatic energy, gained due to charging of the equivalent droplet-dielectric capacitor during electrowetting, also cannot be trivially precluded from the present analysis of electrowetting on soft, dielectric elastomer films.

*3.1 Dielectric film surface deformation by an electrowetted sessile droplet*

The EWOD experimental setup used in the present work is similar to the EWOD systems used in majority of the electrowetting applications[1]. Such EWOD systems involve manipulations of droplets, with characteristic length scales of the order of few hundreds of microns to a few millimetres, on thin dielectric films, with thickness $O(h) \sim 1\,\mu\text{m}$. Hence, for these systems $O(h/r_c^{eq})$ is quite small $\sim 10^{-3} - 10^{-2}$. Consequently, for the present electrowetted sessile droplet and dielectric system, and for most of the EWOD systems in general, a two-dimensional analysis, within the framework of a Cartesian co-ordinate system, will provide a good estimation of the underlying film surface deformation, as evidenced for dimensionally similar soft wetting problems without electrical effects[17,19]. Furthermore, in case of droplet spreading and wetting on soft elastomer films (like Sylgard 184 films) and gels, the resulting film surface deformation is usually analysed following the linear elasticity theory, without any loss of generality[12,13,17,19,16,24–29]. In accordance with the established literature on soft wetting without electrical effects, we will also analyze the deformation of the soft, dielectric elastomer film surface, due to an electrowetted sessile droplet, by considering the dielectric film as an isotropic, linear elastic medium.

Based on the aforementioned assumptions, and on the co-ordinate system shown in the blow-up schematic in Fig. 1(b), the linear elastic constitutive relation for the dielectric films can be written as

$$\sigma_{ij} = \frac{E}{1+\nu}\left[\frac{\nu}{1-2\nu}\left(\frac{\partial u_r}{\partial x_r} + \frac{\partial u_z}{\partial x_z}\right)\delta_{ij} + \frac{1}{2}\left(\frac{\partial u_i}{\partial x_j} + \frac{\partial u_j}{\partial x_i}\right)\right] \quad (1)$$

where $\sigma_{ij}$ is the total stress, $\vec{u} = u_r \hat{r} + u_z \hat{z}$ is the displacement vector, $u_r$ is the displacement component along the $x_r$-direction (unit vector along this direction is $\hat{r}$), $u_z$ is the displacement component along the $x_z$-direction (unit vector along this direction is $\hat{z}$), $\delta_{ij}$ is the Kronecker delta, and $\nu$ is the Poisson's ratio. Here, $i,j = 1,2$, where $i,j = 1$ represents the r-direction and $i,j = 2$ represents the z-direction. Incorporating Eq. (1) into the Navier's equation under equilibrium condition: $\frac{\partial \sigma_{ij}}{\partial x_j} = 0$, the governing equations for the displacements $(u_r, u_z)$ at any location $(x_r, x_z)$ in the dielectric film can be written as

$$(1-2\nu)\left(\frac{\partial^2 u_r}{\partial x_r^2} + \frac{\partial^2 u_r}{\partial x_z^2}\right) + \frac{\partial}{\partial x_r}\left(\frac{\partial u_r}{\partial x_r} + \frac{\partial u_z}{\partial x_z}\right) = 0 \quad (2a)$$

$$(1-2\nu)\left(\frac{\partial^2 u_z}{\partial x_r^2} + \frac{\partial^2 u_z}{\partial x_z^2}\right) + \frac{\partial}{\partial x_z}\left(\frac{\partial u_r}{\partial x_r} + \frac{\partial u_z}{\partial x_z}\right) = 0 \quad (2b)$$

The boundary conditions for this system of equations can be written as

At the interface where the dielectric elastomer film is attached to the rigid ITO coated glass slide, displacements must be zero:

$$u_r(x_r, x_z = 0) = 0; \quad u_z(x_r, x_z = 0) = 0 \tag{3a}$$

At the interface of the electrowetted sessile droplet and the dielectric elastomer film, the stress conditions are:

$$\sigma_{zz}(x_z = h) = \sigma_{zz}^{ext} + \sigma_{zz}^{S} \quad \text{(Normal stress condition)} \tag{3b}$$

$$\sigma_{rz}(x_z = h) = 0 \quad \text{(Tangential stress condition)} \tag{3c}$$

Here $\sigma_{zz}^{ext}$ is the external normal stress acting on the dielectric film surface due to the electrowetted sessile droplet; the exact expression for $\sigma_{zz}^{ext}$ will be discussed later on. Furthermore, $\sigma_{zz}^{S}$ is the inherent internal contribution to the normal stress condition at the droplet-dielectric film interface, due to the surface stress $(\Gamma_S)$ of the solid dielectric film[17,19,30]: $\sigma_{zz}^{S} = \Gamma_S \kappa = \Gamma_S \left. \frac{\partial^2 u_z}{\partial x_r^2} \right|_{x_z = h}$, where $\kappa$ is the curvature of the deformed film surface along $\hat{z}$. $\Gamma_S$ physically represents the reversible work, per unit area, associated with the variation in the excess free energy of the film surface due to creation of new area on the film surface, due to elastic stretching[17,19,30]. Mathematically, the inclusion of the contribution of $\Gamma_S$ provides a better prediction of the film displacement solution in the immediate vicinity of the droplet contact line, by circumventing the strain singularity[17,19]. Finally, for the axisymmetric, droplet wetting under consideration, the residual tangential stress on the dielectric film surface, due to the electrowetted sessile droplet, is zero.

Since the governing equations- Eq. (2a) and Eq. (2b), are linear, the total stresses on the dielectric film surface $(x_z = h)$ can be related to the displacements at any height $x_z$ in the film, in Fourier space, as[17,19,31]

$$\sigma_{iz}(k, h) = Q_{ji}(k, h, x_z) u_j(x_z) \tag{4}$$

where $k$ is the wave number, and $Q_{ji}$ are the components of the stiffness matrix. Eq. (4) can be rewritten as[17,19]

$$\sigma_{iz}^{ext}(k,h) = Q_{ji}(k,h,x_z)u_j(x_z) - \sigma_{zz}^{S}\delta_{iz}$$
$$\Rightarrow \sigma_{iz}^{ext}(k,h) = Q_{ji}^{mod}(k,h,x_z)u_j(x_z) \quad (5)$$

where $Q_{ji}^{mod}$ are the components of the modified stiffness matrix, which incorporates the consequence of the free energy penalty (embodied by $\Gamma_S$) due to creation of new area on the dielectric film surface, due to elastic deformation. Eq. (5) can be expressed in terms of the Fourier transforms of the stress and displacement components as

$$\begin{bmatrix} \hat{\sigma}_{rz}^{ext} \\ \hat{\sigma}_{zz}^{ext} \end{bmatrix} = \begin{bmatrix} Q_{rr}^{mod} & Q_{zr}^{mod} \\ Q_{rz}^{mod} & Q_{zz}^{mod} \end{bmatrix} \begin{bmatrix} \hat{u}_r \\ \hat{u}_z \end{bmatrix} \quad (6)$$

where $\hat{u}_r = \mathfrak{F}[u_r]$, $\hat{u}_z = \mathfrak{F}[u_z]$, $\hat{\sigma}_{rz}^{ext} = \mathfrak{F}[\sigma_{rz}^{ext}]$, and $\hat{\sigma}_{zz}^{ext} = \mathfrak{F}[\sigma_{zz}^{ext}]$ are the Fourier transforms $\{\mathfrak{F}[\ ]\}$ of the corresponding quantities. Furthermore, on expressing Eq. (2a) and Eq. (2b) in terms of the Fourier transforms of the film displacement components, and then, on subsequently solving the resulting equations with the help of the boundary condition presented in Eq. (3a), we have

$$\begin{bmatrix} \hat{u}_r \\ \hat{u}_z \end{bmatrix} = [N] \begin{bmatrix} \dfrac{\partial \hat{u}_r}{\partial x_z}\bigg|_{x_z=0} \\ \dfrac{\partial \hat{u}_z}{\partial x_z}\bigg|_{x_z=0} \end{bmatrix} \quad (7a)$$

where

$$[N] = \begin{bmatrix} \dfrac{(3-4\nu)\sinh(kx_z) + kx_z \cosh(kx_z)}{4k(1-\nu)} & -i\dfrac{x_z \sinh(kx_z)}{2(1-2\nu)} \\ -i\dfrac{x_z \sinh(kx_z)}{4(1-\nu)} & \dfrac{(3-4\nu)\sinh(kx_z) - kx_z \cosh(kx_z)}{2k(1-2\nu)} \end{bmatrix} \quad (7b)$$

Similarly, on expressing the Fourier transforms of the components of the stress, at the dielectric film surface, in terms of $\hat{u}_r$ and $\hat{u}_z$ using Eq. (1), and then, on using Eq. (7a), it can be shown by comparing the resulting expressions for $\hat{\sigma}_{rz}^{ext}$ and $\hat{\sigma}_{zz}^{ext}$ with Eq. (6) that[17,19]

$$\left[Q^{mod}(k,h,x_z)\right] = \begin{bmatrix} Q_{rr}^{mod} & Q_{zr}^{mod} \\ Q_{rz}^{mod} & Q_{zz}^{mod} \end{bmatrix} = \begin{bmatrix} \dfrac{E}{2(1+v)} & 0 \\ 0 & \dfrac{E}{(1+v)(1-2v)} \end{bmatrix} [P][N]^{-1} + \begin{bmatrix} 0 & 0 \\ 0 & \Gamma_S k^2 \end{bmatrix} \quad (8a)$$

where

$$[P] = \begin{bmatrix} 0 & ik \\ ikv & 0 \end{bmatrix}[N] + \begin{bmatrix} 1 & 0 \\ 0 & 1-v \end{bmatrix} \frac{\partial}{\partial x_z}[N] \quad (8b)$$

Now, from Eq. (6) we have

$$\begin{bmatrix} \hat{u}_r \\ \hat{u}_z \end{bmatrix} = \begin{bmatrix} Q_{rr}^{mod} & Q_{zr}^{mod} \\ Q_{rz}^{mod} & Q_{zz}^{mod} \end{bmatrix}^{-1} \begin{bmatrix} \hat{\sigma}_{rz}^{ext} \\ \hat{\sigma}_{zz}^{ext} \end{bmatrix} \quad (9)$$

Here, the elements of the inverse of the modified stiffness matrix can be evaluated by first determining the modified stiffness matrix $\left[Q^{mod}(k,h,x_z=h)\right]$ using Eq. (8a), Eq. (8b), and Eq. (7b) (with $x_z = h$), and then taking the inverse of the resulting stiffness matrix. It must be noted here that according to Eq. (3c): $\sigma_{rz}(x_z = h) = 0 \Rightarrow \hat{\sigma}_{rz} = 0$. So finally, taking the inverse Fourier transform and using Eq. (9), the in-plane $(u_r)$ and out-of-plane $(u_z)$ components of the dielectric film surface $(x_z = h)$ displacement, due to the electrowetted sessile droplet, can be written as

$$u_r(x_z = h) = \sqrt{\frac{2}{\pi}} \int_0^\infty \left(Q_{zr}^{mod}(k,h,x_z=h)\right)^{-1} \hat{\sigma}_{zz}^{ext} i \sin(kx_r) \, dk \quad (10a)$$

$$u_z(x_z = h) = \sqrt{\frac{2}{\pi}} \int_0^\infty \left(Q_{zz}^{mod}(k,h,x_z=h)\right)^{-1} \hat{\sigma}_{zz}^{ext} \cos(kx_r) \, dk \quad (10b)$$

Here, the involved elements of the inverse of the modified stiffness matrix are

$$\left(Q_{zr}^{mod}(k,h,x_z=h)\right)^{-1}$$

$$=-\frac{i(1+\nu)h}{E}\left[\frac{\dfrac{\left(-3+10\nu-8\nu^2-2\bar{k}^2\right)+\left(3-10\nu+8\nu^2\right)\cosh(2\bar{k})}{(3-4\nu)\sinh(2\bar{k})-2\bar{k}}}{\dfrac{\left(5-12\nu+8\nu^2+2\bar{k}^2\right)+(3-4\nu)\cosh(2\bar{k})}{(3-4\nu)\sinh(2\bar{k})-2\bar{k}}\bar{k}+\dfrac{2(1-\nu^2)\Gamma_S\bar{k}^2}{Eh}}\right] \quad (11a)$$

$$\left(Q_{zz}^{mod}(k,h,x_z=h)\right)^{-1}$$

$$=\frac{2(1-\nu^2)h}{E}\left[\frac{1}{\dfrac{\left(5-12\nu+8\nu^2+2\bar{k}^2\right)+(3-4\nu)\cosh(2\bar{k})}{(3-4\nu)\sinh(2\bar{k})-2\bar{k}}\bar{k}+\dfrac{2(1-\nu^2)\Gamma_S\bar{k}^2}{Eh}}\right] \quad (11b)$$

where $\bar{k}=kh$.

The normal stress on the dielectric film surface, due to the electrowetting of a perfectly conducting sessile droplet on it, can be written as

$$\sigma_{zz}^{ext}=\gamma\sin\theta_{eq}\delta\left(x_r-r_c^{eq}\right)-\sigma_{MS}\,\mathrm{H}\left(r_c^{eq}-x_r\right) \quad (12)$$

where $\delta\left(x_r-r_c^{eq}\right)$ is the Dirac delta function and $\mathrm{H}\left(r_c^{eq}-x_r\right)$ is the Heaviside step function. The first term on the right hand side of Eq. (12) represents the total residual force on the dielectric surface at the droplet contact line, due to the liquid surface tension. For an electrowetted sessile droplet on a thin dielectric film, with $O(h)\sim O(\Gamma_S/E), O(\gamma/E)$, the force on the dielectric surface due to the droplet TPCL can be approximated by considering the macroscopic equilibrium droplet contact angle. Here, $\gamma/E$ is the classical elastocapillary length scale[17–19]. The second term on the RHS of Eq. (12) represents the stress distribution on the dielectric film surface, underneath the droplet, due to the effect of the Maxwell stress distribution at the dielectric-droplet interface under an applied electrical voltage[1,2,32]. It must be noted that the effect of the Laplace pressure inside the droplet is neglected here because generally for large droplets, as considered here- $O\left(h/r_c^{eq}\right)\sim 10^{-3}-10^{-2}$, the effect of the Laplace pressure on the film surface deformation is observed to be negligible[17,19,16,33]. The electrically induced stress distribution $(\sigma_{MS})$ on the dielectric elastomer film surface underneath the wetted droplet, at a definite magnitude of $V$, can be written as[1,32]

$$\sigma_{MS} = \frac{\epsilon_0 \epsilon_r(E) |\vec{E}_f|^2}{2} \qquad (13)$$

where $\epsilon_r(E)$ is the film elasticity dependent dielectric constant of the elastomer film, and $|\vec{E}_f|^2 \approx \left(\frac{V}{h}\right)^2$. On incorporating Eq. (13) into Eq. (12), $\sigma_{zz}^{ext}$ can be written as

$$\sigma_{zz}^{ext} = \gamma \sin\theta_{eq} \delta(x_r - r_c^{eq}) - \frac{\epsilon_0 \epsilon_r(E) |\vec{E}_f|^2}{2} H(r_c^{eq} - x_r) \qquad (14)$$

Taking the Fourier transform of Eq. (14), and incorporating it into Eq. (10a) and Eq. (10b), the components of the film surface displacement, due to the electrowetted sessile droplet, can be finally written as

$$u_r(x, x_z = h) = \frac{\gamma \sin\theta_{eq}(1+\nu)}{\pi E} \int_0^\infty (\bar{Q}_{zr}^{mod})^{-1} \sin\left(\bar{k}\frac{x}{h}\right) d\bar{k}$$

$$- \frac{(1+\nu)}{E} \frac{\epsilon_0 \epsilon_r(E) |\vec{E}_f|^2}{2} \int_0^\infty (\bar{Q}_{zr}^{mod})^{-1} \frac{h}{\pi} \frac{\cos\left(\bar{k}\frac{x}{h}\right)}{\bar{k}} d\bar{k} \qquad (15a)$$

$$u_z(x, x_z = h) = \frac{2\gamma \sin\theta_{eq}(1-\nu^2)}{\pi E} \int_0^\infty (\bar{Q}_{zz}^{mod})^{-1} \cos\left(\bar{k}\frac{x}{h}\right) d\bar{k}$$

$$+ \frac{2(1-\nu^2)}{E} \frac{\epsilon_0 \epsilon_r(E) |\vec{E}_f|^2}{2} \int_0^\infty (\bar{Q}_{zz}^{mod})^{-1} \left[\frac{h}{\pi} \frac{\sin\left(\bar{k}\frac{x}{h}\right)}{\bar{k}} - \delta\left(\frac{\bar{k}}{h}\right) \cos\left(\bar{k}\frac{x}{h}\right)\right] d\bar{k}$$

(15b)

where $x = x_r - r_c^{eq}$ represents a horizontally shifted coordinate at the droplet edge (see the blow-up schematic in Fig. 1(b)), and $(\bar{Q}_{zr}^{mod})^{-1}$ and $(\bar{Q}_{zz}^{mod})^{-1}$ are the quantities within the brackets ([ ]) in Eq. (11a) and Eq. (11b) respectively.

### *3.2 Elastic strain energy stored due to the dielectric film surface deformation*

The elastic strain energy stored due to the deformation of the dielectric film surface, as given by Eq. (15a) and Eq. (15b), is defined as

$$E_{el} = \frac{1}{2} \int_{A_{SL}} \sigma_{zz}^{ext} \hat{z} \cdot \left[ u_r(x, x_z = h)\hat{r} + u_z(x, x_z = h)\hat{z} \right] dA_{SL}$$

$$\Rightarrow E_{el} = \frac{1}{2}\left(2\pi r_c^{eq} \gamma \sin\theta_{eq}\right) u_z(x = 0, x_z = h) - \frac{\epsilon_0 \epsilon_r(E) |\vec{E}_f|^2}{4} \int_{A_{SL}} u_z(x, x_z = h) \, dA_{SL} \quad (16)$$

The first term on the RHS of Eq. (16) represents the elastic strain energy stored due to the creation of the wetting ridge about the droplet TPCL, due to the electrically mediated elastocapillary interactions, or electro-elastocapillary interactions, between the electrowetted droplet and the dielectric film. The second term on the RHS of Eq. (16) represents the elastic energy stored due to the entire deformation of the dielectric film underneath the droplet. Since for the droplet and the dielectric film system $\left(O(h/r_c^{eq}) \sim 10^{-3} - 10^{-2}\right)$ considered here, $u_z \to 0$ inside the droplet for the soft dielectric elastomer film (as will be clearly shown during the discussions on the film surface displacement profiles in section 4.2 of this paper), the second term on the RHS can be neglected without loss of generality. So, the total stored elastic energy given by Eq. (16) can be re-written, using the expression for $u_z$ in Eq. (15b), as

$$E_{el} \approx \frac{2(1-\nu^2)}{E} r_c^{eq} \gamma^2 \sin^2\theta_{eq} I_1(x=0) + \frac{2(1-\nu^2)}{E} \pi r_c^{eq} \gamma \sin\theta_{eq} \frac{\epsilon_0 \epsilon_r(E)|\vec{E}_f|^2}{2} I_2(x=0)$$

(17)

where $I_1(x=0)$ and $I_2(x=0)$ are the values of the first and second integrals respectively, appearing on the RHS of Eq. (15b), at $x = 0$.

### *3.3 Total free energy of the droplet-dielectric system in electrowetting-on-dielectric configuration, considering the elastic strain energy*

The total free energy of the conducting droplet and dielectric film system in the EWOD configuration (see Fig. 1(b)), considering the fact that the length scale for the dielectric film deformation ($O(u_z) \sim 10^{-6}$ m) is very small compared to the droplet length scale ($O(r_c^{eq}) \sim 10^{-3}$ m), can be written as

$$F \approx E_{surf} + E_e^{net} + E_{el} - \lambda \Lambda$$
$$\approx 2\pi R_d^2 (1 - \cos\theta_{eq})\gamma + \pi R_d^2 \sin^2\theta_{eq} \left[ (\gamma_{SL,eff}^0 - \gamma_{SV}) - \frac{\epsilon_0 \epsilon_r(E)}{2h} V^2 \right] + E_{el} - \lambda\Lambda \quad (18)$$

where $E_{surf}$ is the contribution of the surface energy, $E_e^{net} = \frac{\epsilon_0 \epsilon_r(E)}{2h} V^2$ is the net electrostatic contribution to the free energy inherent to the electrowetting phenomenon[1,2], $E_{el}$ is the elastic energy contribution to the free energy given by Eq. (17), $\lambda$ is the Lagrange multiplier which enforces the constant volume $\left( \Lambda = \pi R_d^3 \left( \frac{2}{3} - \frac{3\cos\theta_{eq}}{4} + \frac{\cos 3\theta_{eq}}{12} \right) \right)$ constraint for the electrowetted sessile droplet, and $R_d$ is the droplet radius which is related to $r_c^{eq}$ as $r_c^{eq} = R_d \sin\theta_{eq}$. Here, $\gamma_{SL,eff}^0$ is the effective solid-liquid interfacial tension without any electrical effects, and $\gamma_{SV}$ is the solid-vapour interfacial tension. Furthermore, here we have considered the alteration in $E_e^{net}$, at a definite value of the applied electrical voltage, due to the variation in film elasticity, by considering an elasticity dependent dielectric constant of the dielectric elastomer film $(\epsilon_r(E))$, as explained previously.

### 3.4 Minimization of the total free energy

The total free energy $(F)$ of the system, as given by Eq. (18), is minimized here following the classical methodology[1,2], to obtain the following equation:

$$\left[ \cos\theta_{eq} + \left( \frac{\gamma_{SL,eff}^0 - \gamma_{SV}}{\gamma} - \frac{\epsilon_0 \epsilon_r(E)}{2h\gamma} V^2 \right) \right] (1 - \cos\theta_{eq})^2$$
$$- \frac{1.5 I_1(x=0)}{2\pi} \frac{\gamma}{E r_c^{eq}} (1 - \cos^2\theta_{eq})(2\cos^4\theta_{eq} - 7\cos^2\theta_{eq} + 6\cos\theta_{eq} - 1)$$
$$- \frac{1.5 I_2(x=0)}{2} \frac{\epsilon_0 \epsilon_r(E) |\vec{E}_f|^2}{2 E r_c^{eq}} (1 - \cos^2\theta_{eq})^{1/2} (\cos^4\theta_{eq} - 4\cos^2\theta_{eq} + 4\cos\theta_{eq} - 1) = 0$$

(19)

Here we have considered $v = 0.5$ in accordance with the established literature on silicone based elastomers in general, and on Sylgard 184 in specific[11–13,17,19,16,31]. Eq. (19) can be solved numerically to obtain the cosine of the macroscopic equilibrium droplet contact angle during electrowetting of a conducting droplet on dielectric films of varying elasticity. It must be noted here that for an apparently rigid dielectric film, Eq. (18) can be minimized, without considering the effects of variation of $E$ (i.e. without including $E_{el}$ and $\epsilon_r = \text{func}(E)$), to get back the classical Lippmann-Young equation[1,2]:

$$\cos\theta_{eq} = \cos\theta_{eq}^0 + \frac{\epsilon_0 \epsilon_r}{2h\gamma} V^2 \qquad (20)$$

## 4 Results and discussions

### 4.1 *Electrowetting characteristics of sessile drops on dielectric elastomer films of different elasticities*

The electrowetting-on-dielectric or electrowetting characteristics of a conducting sessile droplet on dielectric elastomer films of different elasticity, as represented here by the different values of $E$, are shown in Fig. 2. The electrowetting characteristics are depicted here by the variation of the cosine of the macroscopic equilibrium droplet contact angle $(\cos\theta_{eq})$ with the square of the non-dimensional applied electrical voltage $(\bar{V}^2)$ (see Fig. 2(a)), and by the variation of the non-dimensional equilibrium droplet contact radius $(\bar{r}_c^{eq})$ with $\bar{V}$ (see Fig. 2(b)). In Fig. 2(a), the variation of $\cos\theta_{eq}$ is presented in the form of the variation of the quantity $(\cos\theta_{eq} - \cos\theta_{eq}^0)$ for easier interpretation of the measured data, where $\theta_{eq}^0$ is the macroscopic equilibrium droplet contact angle obtained without any applied electrical voltage. Furthermore, here $\bar{r}_c^{eq} = r_c^{eq}/r_c^0$ and $\bar{V} = V/V_{max}$, where $r_c^0$ is the equilibrium droplet contact radius at zero applied electrical voltage, and $V_{max}(=150$ V) is the maximum applied electrical voltage as discussed in sub-section 2.2.

At a definite magnitude of the applied electrical voltage, the experimentally measured reduction in $\theta_{eq}$ compared to $\theta_{eq}^0$, due to the electrowetting phenomenon, is more for the droplet on the 10:1 Sylgard 184 dielectric film ($E = 1.5$ MPa), than that measured for the droplet on the 50:1 Sylgard 184 dielectric film ($E = 0.02$ MPa) (see Fig. 2(a)). So, the extent of reduction in $\theta_{eq}$ with increasing magnitude of the applied electrical voltage decreases with decreasing elasticity or increasing softness of the dielectric film (see Fig. 2(a)). Furthermore, at a definite value of the applied electrical voltage, the value of $\bar{r}_c^{eq}$ is less for the droplet on the dielectric film with $E = 0.02$ MPa, than that observed for the droplet on the dielectric film with $E = 1.5$ MPa (see Fig. 2(b)). In essence, it can be concluded here that the extent of electrowetting of a sessile droplet reduces (i.e. $\theta_{eq}$ increases and $r_c^{eq}$ decreases) with decreasing value of the Young's modulus $(E)$ of the underlying dielectric film (see Fig. 2(a) and Fig. 2(b)). Moreover, the variation of $\theta_{eq}$ with the applied electrical voltage, for the droplet on the dielectric film with $E = 1.5$ MPa, is well described by the classical Lippmann-Young equation, i.e. Eq. (20) (see Fig. 2(a)). This proves that the droplet electrowetting on the dielectric elastomer film with $E = 1.5$ MPa adheres to the classical description of the electrowetting phenomenon, which does not consider the consequences of the variation in $E$ beyond the 'rigid' domain. Furthermore, it can be now unequivocally concluded that the dielectric elastomer film, with $E = 1.5$ MPa, indeed behaves as an apparently rigid dielectric film, as mentioned in sub-section 2.1. Fig. 2(a) also indicates that the Lippmann-Young equation (Eq. (20)) fails to address the variation of $\theta_{eq}$ with the applied electrical voltage for the droplet on the dielectric elastomer film with a lower value of the Young's modulus, i.e. $E = 0.02$ MPa. This in turn implies that the electrowetting on the soft, deformable dielectric film ($E = 0.02$ MPa) deviate from the classical description of the electrowetting phenomenon, as established in the existing literature.

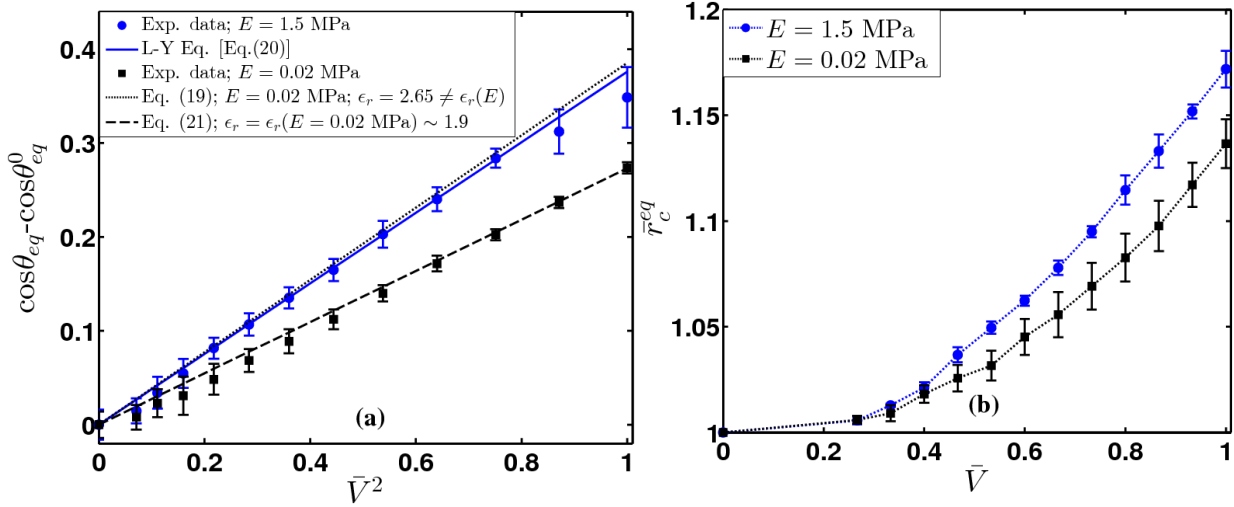

**Fig. 2: (a)** Variations of the cosine of the macroscopic equilibrium droplet contact angle $\left(\cos\theta_{eq}\right)$ with the increasing magnitude of the square of the non-dimensional applied electrical voltage $\left(\bar{V}^2\right)$, for the 100 mM KCl solution droplet, on the dielectric elastomer films having different values of the Young's modulus- $E = 1.5$ MPa and $E = 0.02$ MPa. The Lippmann-Young (L-Y) equation (Eq. (20)) suitably describes the droplet electrowetting on the apparently rigid dielectric film ($E = 1.5$ MPa), but fails to describe the same on the soft, deformable dielectric film ($E = 0.02$ MPa). The electrowetting behaviour on the soft, deformable dielectric film ($E = 0.02$ MPa) is well addressed only by the modified Lippmann-Young equation (Eq. (21)), which takes into consideration the effective dependence of the dielectric constant $(\epsilon_r)$ on the elasticity, for a general dielectric elastomer film. The elastic strain energy due to the deformation of the soft film surface does not influence the macroscopic equilibrium droplet configuration on the soft dielectric film, under an applied electrical voltage. **(b)** Variations of the non-dimensional equilibrium droplet contact radius $\left(\bar{r}_c^{eq}\right)$ with the increasing magnitude of $\bar{V}$, for the 100 mM KCl solution droplet, on the dielectric elastomer films having different values of $E$.

To understand the physical reason behind the observed deviation of the electrowetting behaviour on the dielectric elastomer film with $E = 0.02$ MPa, from the classical description, we plot in Fig. 2(a) the variation of $\cos\theta_{eq}$ obtained by numerically solving Eq. (19) at different values of $V$, considering $E = 0.02$ MPa. During this solution procedure, we first ignore the corresponding change in the dielectric constant $(\epsilon_r)$ due to the reduction of $E$, from 1.5 MPa to 0.02 MPa, for the dielectric elastomer. Hence, $\epsilon_r$ is still considered to be equal to 2.65, which is ideally the value of the dielectric constant for the apparently rigid Sylgard 184 dielectric film. This methodology is adopted here to first highlight upon the influence (if any) of the additional elastic energy, stored due to the deformation of the soft dielectric film surface by the electrowetted sessile droplet, on the final macroscopic equilibrium configuration of the droplet characterized by $\theta_{eq}$. Interestingly, Fig. 2(a) clearly shows that the solutions of Eq. (19), under the aforementioned conditions ($E = 0.02$; $\epsilon_r = 2.65 \neq \epsilon_r(E)$), also fail to describe the variation of $\theta_{eq}$ with the applied electrical voltage on the dielectric elastomer film with $E = 0.02$ MPa (see Fig. 2(a)). So, for a conducting sessile droplet on a thin dielectric film, where $O(h/r_c^{eq}) \sim 10^{-3} - 10^{-2}$, the deformation of the soft dielectric film surface does not alter the value of the macroscopic equilibrium droplet contact angle under an applied electrical voltage, from that observed on the corresponding apparently rigid dielectric film under identical conditions (see Fig. 2(a)). The practically insignificant difference that is observed is due to the slight difference in $h$ for the films with $E = 1.5$ MPa and $E = 0.02$ MPa. It must be noted here that for solving Eq. (19), we have chosen the order of magnitude of the surface stress ($\Gamma_S = 0.1$ N/m), for the soft dielectric film, in accordance with the established literature[17,19]. Based on the above discussion, it can be concluded that the additional elastic energy is not responsible for the observed alteration of the macroscopic equilibrium configuration of the electrowetted sessile droplet on the soft dielectric elastomer film, from that observed on the apparently rigid dielectric elastomer film under identical conditions. So, on neglecting the elastic energy contribution to the free energy, but on considering the influence of the effective dependence of $\epsilon_r$ on $E$, Eq. (19) can be rewritten as

$$\cos\theta_{eq} \approx \cos\theta_{eq}^0 + \frac{\epsilon_0 \epsilon_r(E)}{2h\gamma} V^2 \tag{21}$$

On considering $\epsilon_r = \epsilon_r(E = 0.02 \text{ MPa}) \sim 1.9$ (and not equal to 2.65), Eq. (21) suitably describes the variation of $\theta_{eq}$, with the applied electrical voltage, for the conducting sessile droplet on the dielectric elastomer film with $E = 0.02$ MPa (see Fig. 2(a)). It must be noted here that $\epsilon_r \sim 1.9$ conforms to the value of the dielectric constant for Sylgard 184 with low degree of polymerization[23], and is not any *ad hoc* value chosen to fit the experimental data. So, the effective dependence of $\epsilon_r$ on $E$, due to their mutual dependence on the degree of polymerization for a general dielectric elastomer, primarily dictates the electrically controlled variation of $\theta_{eq}$ on dielectric elastomer films of different elasticity. For the soft, deformable dielectric film, the reduction in $\epsilon_r$ (from $\sim 2.65$ to $\sim 1.9$), accompanying the reduction in $E$ from the 'apparently rigid' domain to the 'soft' domain, reduces the net electrostatic contribution $\left(E_e^{net} = \frac{\epsilon_0 \epsilon_r(E)}{2h} V^2\right)$ to the total free energy of the EWOD system. This reduction in the electrostatic energy gained due to the charging of the equivalent dielectric-droplet capacitor, for the conducting droplet and the soft dielectric film, effectively manifests in reduced extent of electrowetting of the sessile drop on it. This is in accordance with the thermodynamic understanding of the electrowetting mechanism[1]. At a definite magnitude of the applied electrical voltage, the reduced extent of electrowetting of the sessile drop on the soft dielectric elastomer film, with $E = 0.02$ MPa, gets reflected by the increase in $\theta_{eq}$ and corresponding decrease in $r_c^{eq}$, as compared to the values of these quantities on the apparently rigid dielectric elastomer film, with $E = 1.5$ MPa. It must be noted here that although Eq. (21) looks similar to the classical Lippmann-Young equation (Eq. (20)), philosophically these are quite different. The classical derivation of the Lippmann-Young equation (Eq. (20)) by the thermodynamic approach does not take in purview the contribution of the elastic energy, due to the deformation of the soft dielectric film surface. This elastic energy is initially considered during the derivation of Eq. (21), but has been subsequently neglected since it fails to trigger any macroscopic alteration of the electrowetting behaviour on the soft dielectric film. This fact is represented in Eq. (21) by the use of '$\approx$' instead of '$=$'. More importantly, the classical Lippmann-Young equation (Eq. (20)) does not consider the effective coupling between $\epsilon_r$ and $E$ for general dielectric elastomers. However, it is precisely this classically neglected aspect which is incorporated in Eq. (21) by writing $\epsilon_r$ as $\epsilon_r(E)$. It is this consideration of $\epsilon_r$ as an effective function of $E$ which has been shown here to play the pivotal role in describing the alterations in the macroscopic electrowetting characteristics of a sessile droplet, due to the reduction of the elasticity of the dielectric elastomer film from the 'apparently rigid' to the 'soft' domain. So, under an applied electrical voltage, the final equilibrium configuration of a sessile droplet, as characterized by $\theta_{eq}(V;E)$, is dictated by the combined influences of the magnitude of $V$ and the magnitude of $E$. The dependence of $\theta_{eq}$ on $E$ stems from the effective dependence of $\epsilon_r$ on $E$ for general dielectric elastomers. In reality, for electrowetting on a soft, deformable dielectric film, the underlying deformation of the dielectric film surface does not have any macroscopically perceivable influence on $\theta_{eq}$. However, it must be remembered here that the electrospreading characteristics on soft dielectric films, for geometrically similar droplet and dielectric systems $\left(O\left(h/r_c^{eq}\right) \sim 10^{-3} - 10^{-2}\right)$, are dependent on this underlying dielectric film surface deformation[14]. Electrospreading here refers to the spreading process by which the sessile droplet reaches its new thermodynamic equilibrium configuration, specified by $\theta_{eq}(V;E)$, from its original thermodynamic equilibrium state without any electrical effect, characterized by $\theta_{eq}^0$, on application of an electrical voltage. So, it is important to have a discussion on the actual deformation profiles for the soft dielectric film surface, in order to develop truly in-depth understanding of electrically modulated droplet wetting and spreading mechanisms on soft substrates. Moreover, such deformation of soft films due to electro-elastocapillary interactions, during wetting of sessile drops under an applied electrical voltage, is still unaddressed in the 'soft wetting' literature. The displacement profiles for soft films due to wetting of sessile drops, as percolated in the literature so far, do not take into consideration any electrical effect.

## 4.2 Theoretical profiles for soft dielectric film surface deformed by an electrowetted sessile droplet

The out-of-plane ($u_z$; Eq. (15b)) and in-plane ($u_r$; Eq. (15a)) displacement profiles of the soft dielectric film surface ($E = 0.02$ MPa), due to the forcing imposed by the electrowetted sessile droplet on it (see the blow-up schematic in Fig. 1(b)), are shown in Fig. 3(a) and Fig. 3(b) respectively. The displacement profiles of the soft dielectric film surface are presented here as a function of the non-dimensional distance from the droplet three phase contact line (TPCL) $\left( \bar{x} = \frac{x}{h} = \frac{x_r - r_c^{eq}}{h} \right)$. Moreover, $u_z$ and $u_r$ are non-dimensionalized here by the elastocapillary length scale $\gamma/E$. Furthermore, displacement profiles of the soft dielectric film surface, due to the electrowetting of a sessile drop, are shown in Fig. 3(a) and Fig. 3(b) for different magnitudes of the applied electrical voltage. The applied electrical voltage is also presented here in a non-dimensional form as $\bar{V} = V/V_{max}$.

For the droplet and dielectric film system under consideration $\left( O\left(h/r_c^{eq}\right) \sim 10^{-3} - 10^{-2} \right)$, at $\bar{V} = 0$, the normal wetting of the droplet results in the creation of a symmetric wetting ridge on the soft dielectric film surface, about the droplet TPCL (see Fig. 3(a)). The wetting ridge is flanked by a valley both on the left (within the droplet) and on the right of it (outside the droplet) (see Fig. 3(a)). The height of the wetting ridge on the soft dielectric film surface is- $O(u_z) \sim O(\gamma/E) \sim 10^{-6}$ m. However, on application of the external electrical voltage, the additional stress distribution on the soft dielectric film surface underneath the droplet, due to the Maxwell stress distribution at the dielectric-droplet interface, results in the formation of an asymmetric wetting ridge about the TPCL (see Fig. 3(a)). The increasing magnitude of this electrically induced stress distribution on the soft dielectric film surface, with increasing magnitude of $\bar{V}$, progressively increases the depth of the valley at the base of the wetting ridge within the droplet (see Fig. 3(a) and its inset). Consequently, this results in progressive reduction in the depth of the valley at the base of the wetting ridge outside the droplet, with the increasing magnitude of $\bar{V}$ (see Fig. 3(a) and its inset). Such modification of the out-of-plane displacement of the soft dielectric film surface, due to the electrowetted sessile droplet, results in the corresponding asymmetric $u_z$ profile about the TPCL, at a definite magnitude of $\bar{V}$ (see Fig. 3(a)). However, the effect of the electrically induced stress distribution on the out-of-plane displacement of the soft dielectric film surface vanishes within a distance $\sim O(h)$, from the droplet contact line (see Fig. 3(a)). This is because the Maxwell stress distribution at the dielectric-droplet interface reduces to insignificant values within a distance $\sim O(h)$, from the droplet TPCL[1]. Moreover, $u_z \to 0$ within the droplet beyond this small region adjacent to the TPCL, for all values of $\bar{V}$ including $\bar{V} = 0$. This is because of the fact that for the droplet and dielectric film system under consideration $O(h/r_c^{eq})$ is quite small $(\sim 10^{-3} - 10^{-2})$. It must be noted here that the height of the wetting ridge is independent of the electrically induced stress distribution on the soft dielectric film surface underneath the droplet (see Fig. 3(a)); it is only dictated by the residual vertical force at the droplet contact line due to the liquid surface tension. Hence, the length scale for the wetting ridge on the soft surface remains similar to the classical elastocapillary length scale $\gamma/E$, even under an applied electrical voltage.

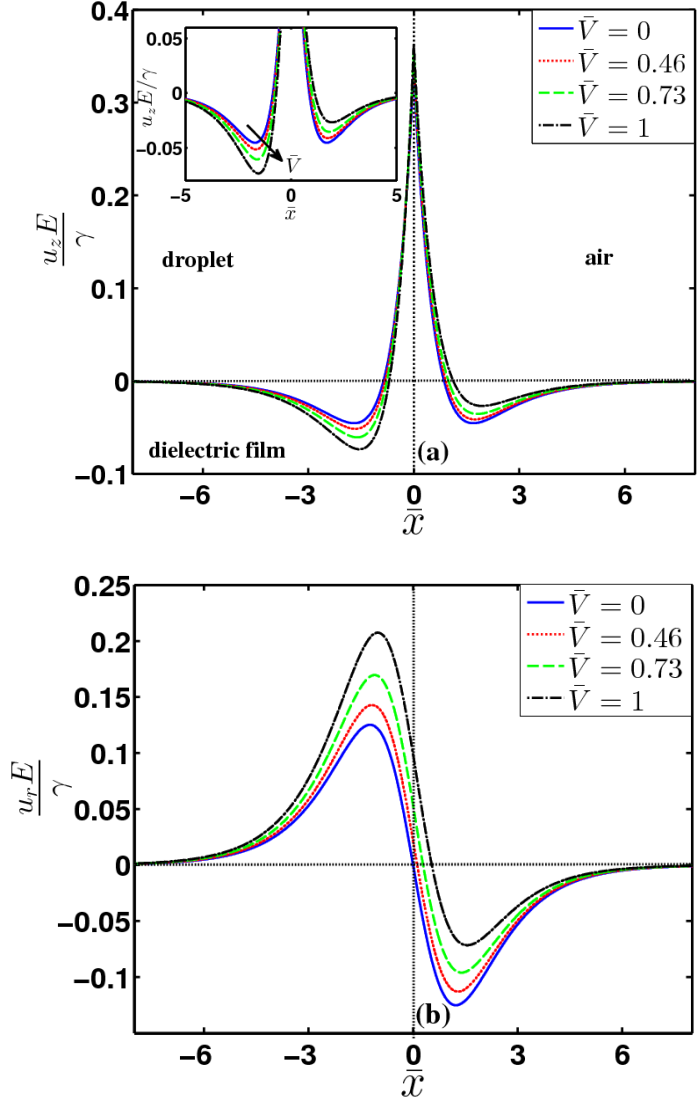

**Fig. 3:** **(a)** The out-of-plane $(u_z)$ displacement profiles, for the soft dielectric film surface, due to the deformation induced by an electrowetted sessile droplet, at different magnitudes of the applied electrical voltage, presented here in a non-dimensional form by $\bar{V} = V/V_{max}$. The inset highlights upon the asymmetry in the profile, induced by the electrically induced stress distribution on the soft dielectric film surface underneath the electrowetted sessile droplet. **(b)** The in-plane $(u_r)$ displacement profiles for the soft dielectric film surface, at different magnitudes of the applied electrical voltage. $u_z$ and $u_r$ are non-dimensionalized here by the elastocapillary length scale $\gamma/E$. $\bar{x} = \dfrac{x_r - r_c^{eq}}{h}$ presents the non-dimensional distance from the droplet three phase contact line (TPCL).

At $\bar{V} = 0$, the in-plane or horizontal displacement of the soft dielectric film surface also shows a symmetric profile about the TPCL (see Fig. 3(b)). The in-plane displacement increases towards the droplet contact line, over the region corresponding to the valley in the $u_z$ profile (compare Fig. 3(a) and Fig. 3(b)). However, $u_r$ starts decaying very close to the TPCL, once the wetting ridge starts taking shape (compare Fig. 3(a) and Fig. 3(b)). For $\bar{V} = 0$, $u_r = 0$ at the droplet contact line (see Fig. 3(b)). In case of the in-plane displacement also, the application of the external electrical voltage results in an asymmetric $u_r$ profile, about the TPCL, for the soft dielectric film surface deformation. The increasing strength of the electrically induced stress distribution on the soft dielectric film surface underneath the electrowetted droplet, with increasing magnitude of $\bar{V}$, results in progressively increasing in-plane displacement within the droplet (see Fig. 3(b)). The region over which the in-plane displacement within the droplet progressively increases, corresponds to the valley of increasing depth at the base of the wetting ridge within the droplet, observed in the $u_z$ profile (compare Fig. 3(a) and Fig. 3(b)). The in-plane displacement outside the droplet adjusts accordingly, finally resulting in an asymmetric $u_r$ profile, at a definite magnitude of $\bar{V}$ (see Fig. 3(b)). Hence, at a definite magnitude of $\bar{V}$, $u_r$ is no longer zero at the droplet contact line (see Fig. 3(b)). Moreover, the effect of the electrically induced stress distribution underneath the electrowetted droplet, on the $u_r$ profile, also tends to vanish within a distance $\sim O(h)$ from the droplet contact line (see Fig. 3(b)). This is because the Maxwell stress distribution tends to vanish over this length scale from the droplet contact line, as mentioned previously. Finally, just like $u_z$, $u_r \to 0$ within the droplet beyond the small region adjacent to the TPCL, for all values of $\bar{V}$ including $\bar{V} = 0$, since $O\left(h/r_c^{eq}\right) \sim 10^{-3} - 10^{-2}$ (see Fig. 3(b)).

## 5  Conclusions and scope

The extent of electrowetting of a conducting sessile droplet on a thin dielectric elastomer film is dependent on the elasticity of the dielectric film, contrary to the general perception percolated in the existing literature. At a definite magnitude of the applied electrical voltage, the macroscopic equilibrium droplet contact angle increases, and the equilibrium droplet contact radius decreases, with decreasing elasticity or increasing softness of the dielectric film. To put it succinctly, the extent of electrowetting decreases with decreasing elasticity of the dielectric elastomer film. In this regard, the electrowetting phenomenon on an apparently rigid, dielectric elastomer film is well described by the classical Lippmann-Young equation. However, the established Lippmann-Young paradigm fails to describe the reduced extent of electrowetting on the corresponding soft, deformable dielectric elastomer film. The reduced extent of electrowetting on the soft dielectric film stems from the involved reduction in the dielectric constant of the film, associated with the reduction in the elasticity of the film from the 'apparently rigid' domain to the 'soft' domain. The effective dependence of the dielectric constant on the elasticity, for a general dielectric elastomer film, stems from the mutual dependence of these quantities on the degree of polymerization of the polymeric chain.

We have shown here that the electrowetting phenomenon on dielectric films of varying elasticities can be addressed, on a mesoscopic to macroscopic scale, by a generalized, or modified, Lippmann-Young equation. This modified Lippmann-Young equation incorporates the effective dependence of the dielectric constant on the elasticity of the film. In this regard, it must be noted here that the deformation of the underlying soft dielectric film surface does not influence the macroscopic droplet equilibrium configuration, under an applied electrical voltage; hence, the influence of the same is not apparently reflected in the modified Lippmann-Young equation. However, it must be acknowledged here that the displacement profile of the soft dielectric film surface, due to the wetting of a sessile droplet under an applied electric voltage, is distinctly different from that established for 'soft wetting' without any electrical effects. The electro-elastocapillarity induced deformation of the soft dielectric film surface results in progressively asymmetric displacement profiles about the droplet contact line, with increasing magnitude of the applied electrical voltage. The electro-elastocapillarity induced asymmetric surface deformation profile is primarily due to the influence of the increasing strength of the electrically induced stress distribution on the soft dielectric film, underneath the electrowetted droplet. However, the height of the wetting ridge on the soft dielectric surface remains independent of this additional effect of the electrically induced stress distribution.

It must be noted here that the various aspects of the electrically modulated wetting of sessile drops on soft dielectric films, as discussed here, are valid for droplet-and-dielectric-film systems satisfying the condition $O\left(h/r_c^{eq}\right) \sim 10^{-3} - 10^{-1}$; interestingly, majority of the EWOD systems satisfy this condition. Hence, the present study provides a practically relevant understanding of electrowetting of sessile drops on rheologically tunable dielectric films of varying elasticity. The influence of the elasticity of the dielectric film on the droplet electrowetting can no longer be trivially neglected *a priori*. Interestingly, the elasticity of the underlying dielectric elastomer film provides an additional control over the droplet electrowetting behaviour on it. The understanding of the inter-connection between droplet electrowetting and dielectric film elasticity developed here can be utilized for designing novel technologies involving the yet nascent paradigm of electro-elastocapillarity mediated droplet wetting. These state-of-the-art technological applications may include the development of soft liquid lenses and flexible reflective displays for better configurability and adaptability. The present research findings may be also used to lay the foundation for new bio-medical applications involving soft interfaces, like control of wetting and spreading of bio-fluids in bio-physical processes, and controlled drug delivery. Further tuning can be imposed on droplet manipulation by using optimal electrode arrays, in tandem with dielectric elasticity gradient, to usher in a new generation of electrically controlled digital microfluidic devices, involving dielectric layers with patterned stiffness. Furthermore, the discussion on soft surface deformation profiles due to electro-elastocapillary interactions, under an electrowetted sessile droplet, may be useful for tuning the functioning of dielectric elastomers in engineering applications pertaining to soft active materials.

**References**

1  J. Berthier, *Microdrops and Digital Microfluidics*, 2008, William Andrew, Norwich, NY, USA.

2  F. Mugele and J.-C. Baret, *J. Phys. Condens. Matter*, 2005, **17**, R705.

3  R. Shamai, D. Andelman, B. Berge and R. Hayes, *Soft Matter*, 2008, **4**, 38.

4  F. Mugele, J. C. Baret and D. Steinhauser, *Appl. Phys. Lett.*, 2006, **88**, 204106.

5  G. Manukyan, J. M. Oh, D. van den Ende, R. G. H. Lammertink and F. Mugele, *Phys. Rev. Lett.*, 2011, **106**, 014501.

6  J. Hong and S. J. Lee, *Lab Chip*, 2015, **15**, 900.

7  P. Y. Paik, V. K. Pamula and K. Chakrabarty, *IEEE Des. Test Comput.*, 2008, **25**, 372.

8  K. Choi, A. H. C. Ng, R. Fobel and A. R. Wheeler, *Annu. Rev. Anal. Chem.*, 2012, **5**, 413.

9  M. Piñeirua, J. Bico and B. Roman, *Soft Matter*, 2010, **6**, 4491.

10  L. Chen and E. Bonaccurso, *Adv. Colloid Interface Sci.*, 2014, **210**, 2.

11  R. Pericet-Camara, G. K. Auernhammer, K. Koynov, S. Lorenzoni, R. Raiteri and E. Bonaccurso, *Soft Matter*, 2009, **5**, 3611.

12  M. C. Lopes and E. Bonaccurso, *Soft Matter*, 2012, **8**, 7875.

13  M. C. Lopes and E. Bonaccurso, *Soft Matter*, 2013, **9**, 7942.

14  R. Dey, A. Daga, S. DasGupta and S. Chakraborty, *Appl. Phys. Lett.*, 2015, **107**, 034101.

15  F. Mugele and J. Buehrle, *J. Phys. Condens. Matter*, 2007, **19**, 375112.

16  M. E. R. Shanahan, *J. Phys. D. Appl. Phys.*, 1987, **20**, 945.

17  E. R. Jerison, Y. Xu, L. A. Wilen and E. R. Dufresne, *Phys. Rev. Lett.*, 2011, **106**,

186103.

18  A. Marchand, S. Das, J. H. Snoeijer and B. Andreotti, *Phys. Rev. Lett.*, 2012, **109**, 236101.

19  R. W. Style and E. R. Dufresne, *Soft Matter*, 2012, **8**, 7177.

20  D. Armani, C. Liu and N. Aluru, *Micro Electro Mech. Syst. 1999. MEMS '99; IEEE Int. Conf.*, 1999, 222.

21  L. Chen, G. K. Auernhammer and E. Bonaccurso, *Soft Matter*, 2011, **7**, 9084.

22  T. L. Magila and D. G. LeGrand, *Pol. Engg. Sci.*, 1970, **10**, 349.

23  F. Gubbels, Ch: Silicones in Industrial Applications, *Inorganic Polymers*, 2007, Nova Science Publishers.

24  M. E. R. Shanahan, *J. Phys. D. Appl. Phys.*, 1988, **21**, 981.

25  M. E. R. Shanahan and A. Carre, *Langmuir*, 1994, **10**, 1647.

26  M. Shanahan and A. Carre, *Langmuir*, 1995, **11**, 1396.

27  A. Carre and M. E. R. Shanahan, *Langmuir*, 1995, **11**, 24.

28  D. Long, A. Ajdari and L. Leibler, *Langmuir*, 1996, **12**, 5221.

29  A. Carre, J. –C. Gastel and M. E. R. Shanahan, *Nature*, 1996, **379**, 432.

30  R. C. Cammarata and K. Sieradzki, *Annu. Rev. Mater. Sci.*, 1994, **24**, 215.

31  Y. Xu, W. C. Engl, E. R. Jerison, K. J. Wallenstein, C. Hyland, L. a Wilen and E. R. Dufresne, *Proc. Natl. Acad. Sci. U. S. A.*, 2010, **107**, 14964.

32  F. Mugele, *Soft Matter*, 2009, **5**, 3377.

33  R. W. Style, R. Boltyanskiy, Y. Che, J. S. Wettlaufer, L. A. Wilen and E. R. Dufresne, *Phys. Rev. Lett.*, 2013, **110**, 066103.